\documentclass[12pt]{article}

\begin{document}
\tolerance=5000
\def\be{\begin{equation}}
\def\ee{\end{equation}}
\def\bea{\begin{eqnarray}}
\def\eea{\end{eqnarray}}
\def\nn{\nonumber \\}
\def\cF{{\cal F}}
\def\det{{\rm det\,}}
\def\Tr{{\rm Tr\,}}
\def\e{{\rm e}}
\def\etal{{\it et al.}}
\def\erp2{{\rm e}^{2\rho}}
\def\erm2{{\rm e}^{-2\rho}}
\def\er4{{\rm e}^{4\rho}}
\def\etal{{\it et al.}}
\def\gsim{\ ^>\llap{$_\sim$}\ }

\ 

\vskip -2cm

\ \hfill
\begin{minipage}{3.5cm}
NDA-FP-62 \\
May 1999 \\
\end{minipage}

\vfill

\begin{center}

{\bf\large
Curvature Dependence of Running Gauge Coupling and Confinement 
in AdS/CFT Correspondence}

\vfill

{\sc Shin'ichi NOJIRI}\footnote{\scriptsize 
e-mail: nojiri@cc.nda.ac.jp} and
{\sc Sergei D. ODINTSOV$^{\spadesuit}$}\footnote{\scriptsize 
e-mail: odintsov@mail.tomsknet.ru}

\vfill

{\sl Department of Mathematics and Physics \\
National Defence Academy, 
Hashirimizu Yokosuka 239, JAPAN}

\ 

{\sl $\spadesuit$ 
Tomsk Pedagogical University, 634041 Tomsk, RUSSIA \\
}

\ 

\vfill

{\bf abstract}

\end{center}

We construct IIB supergravity (viewed as dilatonic gravity) 
background with non-trivial dilaton and with curved 
four-dimensional space. Such a background  may describe another 
vacuum of maximally supersymmetric Yang-Mills theory or strong 
coupling regime of (non)-supersymmetric gauge theory with 
(power-like) running gauge coupling which depends 
on curvature.   
Curvature dependent quark-antiquark 
potential is calculated where the geometry 
type of hyperbolic (or de Sitter universe) shows (or not) 
the tendency of the confinement. 
Generalization of IIB supergravity background with 
non-constant axion is presented. Quark-antiquark potential being 
again curvature-dependent has a possibility to produce the 
standard area law for large separations. 

\vfill

\noindent
PACS: 04.65.+e, 11.15.-q

\newpage

\section{Introduction}

AdS/CFT correspondence \cite{1} may provide new insights to 
the understanding of non-perturbative QCD. For example, 
in frames of Type 0 String Theory the attempts\cite{2} have 
been done to reproduce such well-known QCD effects as 
running gauge coupling and possibly confinement. 
It is among the first problems to get the description of 
well-known QCD phenomena from bulk/boundary correspondence.

In another approach one can consider IIB supergravity (SG) 
vacuum which describes the strong coupling regime of a 
non-supersymmetric gauge theory. 
This can be achieved by the consideration of deformed IIB SG 
vacuums, for an example, with non-constant dilaton which breaks 
conformal invariance and supersymmetry (SUSY) of boundary 
supersymmetric Yang-Mills (YM) theory. Such a background will 
be the perturbation of ${\rm AdS}_5\times{\rm S}_5$ vacuum. 
The background of such a sort (with non-trivial dilaton)
which interpolates between AdS (UV) and flat space with 
singular dilaton (IR) has been found in ref.\cite{3} where 
also conformal dimensions for (dilaton coupled) scalar have 
been found.

This solution of IIB SG \cite{3} has been used in ref.\cite{4} 
with the interpretation of it as the one describing the running 
gauge coupling (via exponent of dilaton). It has been shown that 
running gauge coupling has a power law behavior with ultraviolet 
(UV) stable 
fixed point and quark-antiquark potential\cite{5} has been 
calculated. QCD properties of such a background have been 
discussed in detail in refs.\cite{6}. Modifications of IIB SG 
solution with non-constant dilaton \cite{3} due to presence of 
axion\cite{7}, constant self-dual vector\cite{8} or
world volume scalar \cite{9} give further proof of the possible 
confinement and asymptotic freedom of the boundary non-SUSY gauge 
theory. Unfortunately, situation is very complicated here due to 
double role of IIB SG backgrounds. From one side they may indeed 
correspond to IR gauge theory (deformation of initial 
SUSY YM theory). On the same time such a background may simply 
describe another vacuum of the same maximally supersymmetric 
YM theory with non-zero vacuum expectation value (VEV) 
of some operator. Due to the fact 
that operators corresponding to deformation to another gauge theory 
are not known, it is unclear what is the case under discussion 
(interpretation of SG background). Only some indirect arguments 
as below may be given. As we see these arguments indicate that 
IIB SG background discussed in this work most probably corresponds 
to another vacuum of super YM theory under consideration. Then 
renormalization group (RG) flow is induced in the theory 
via giving a non-zero VEV to some operator.

In the present paper, we continue the study of running dilaton 
and confinement from IIB supergravity backgrounds with non-trivial 
dilaton. We generalize the solution of ref.\cite{3} for non-zero 
curvature of $d$-dimensional space. As a result, IIB supergravity 
background is changed drastically. The running dilaton (gauge 
coupling) depends explicitly on the four-dimensional curvature. 
The structure of quark-antiquark potential is modified. In a sense, 
confinement would become the characteristic of the Universe.

Let us make few remarks on AdS/CFT interpretation of IIB SG 
background. Choosing the coordinates in the asymptotically 
${\rm AdS}_5$ spacetime as 
\be
\label{I1}
ds^2=d\sigma^2 + S(\sigma)\sum_{i,j=0}^{3}\eta_{ij}dx^i dx^j\ ,
\ee
let assume a scalar field $\lambda$, e.g. dilaton, axion or other 
fields, obey the following equation:
\be
\label{I2}
{d^2\lambda \over d\sigma^2} + 4 {d\lambda \over d\sigma}=
M^2 \lambda\ 
\ee
near the boundary. 
Here $M^2$ is the ``mass'' of $\lambda$ and 
$\sigma\rightarrow 0$ corresponds to the boundary of AdS. 
Then $\lambda$ is associated with the operator ${\cal O}_\lambda$ 
with conformal dimension $\Delta = 2+\sqrt{4 + M^2}$. 
The solution of (\ref{I2}) is given by
\be
\label{I3}
\lambda = A \e^{-(4-\Delta)\sigma} + B \e^{-\Delta\sigma}\ .
\ee
The solution corresponding to $A$ is not normalizable but 
the solution to $B$ is normalizable. According to the 
argument in \cite{GPPZ}, the non-normalizable solution would be 
associated with the deformation of the ${\cal N}=4$ theory 
by ${\cal O}_\lambda$ but the normalizable solution would be 
associated with a different vacuum where ${\cal O}_\lambda$ has 
a non-zero vacuum expectation value. The behavior of the dilaton 
found in this paper is normalizable and seems to be associated 
with the dimension 4 operator, say ${\rm tr} F^2$. Then the 
argument in \cite{GPPZ} would indicate that the solution found 
in this paper should correspond to another vacuum of 
${\cal N}=4$ theory. Nevertheless, there might be still 
possibility that the solution corresponds to non-supersymmetric 
gauge theory. Since there occurs the condensation of ${\rm tr} F^2$ 
in the usual non-supersymmetric QCD, however, the solution 
given here would describe some features typical for the 
non-supersymmetric theory.

The situation is even more complicated due to limits of 
validity of dual SG description. 
In order that the classical supergravity description is valid, 
the curvature should be small and the string coupling should be 
also small. If the curvature is large, the $\alpha'$ 
corrections from string theory would appear. In the AdS/CFT 
correspondence, the radius  $R_s$ of the curvature is given by 
\be
\label{L1}
R_s = \left( 4\pi g_s N\right)^{1 \over 4}\ .
\ee
Here $g_s$ is the string coupling and $N$ is the number of the 
coincident D-branes. Therefore we should require
\be
\label{L2}
g_s N \gg 1\ .
\ee
On the other hand, the classical picture works when the string 
coupling is small:
\be
\label{L3}
g_s \ll 1\ .
\ee
In the solution given in this paper, there appears the curvature 
singularity and $g_s$ depends on the coordinates since the 
dilaton is non-trivial. If we concentrate on the behavior near the 
boundary, which is asymptotically AdS and is far from the 
singularity, the solution would be reliable and SG description 
would be trusted.

The work is organized as follows. In the next section we give 
IIB supergravity background with non-constant dilaton 
and non-flat four-dimensional space. Via AdS/CFT it gives 
the curvature dependent (power-like) running gauge coupling 
and quark-antiquark potential where hyperbolic geometry seems to 
support confinement. 
In section 3 we generalize the background of section 2 for the 
case when axion presents. (Curvature dependent) quark-antiquark 
potential is found. It is shown that inflationary Universe 
(de Sitter) with axion might predict confinement. 
Some outlook is given in the last section. Additional solutions of IIB 
supergravity are presented in two Appendixes.

\section{Solution, Running Gauge Coupling 
and Quark-Antiquark Potential}

We start from the following action of dilatonic gravity
in $d+1$ dimensions:
\be
\label{i}
S=-{1 \over 16\pi G}\int d^{d+1}x \sqrt{-G}\left(R - \Lambda 
- - \alpha G^{\mu\nu}\partial_\mu \phi \partial_\nu \phi \right)\ .
\ee
In the following, we assume $\lambda^2\equiv -\Lambda$ 
and $\alpha$ to be positive. 
The action (\ref{i}) is very general. It contains the effective 
action of type IIB string theory. The type IIB supergravity, which 
is the low energy effective action of the type IIB string theory, 
has a vacuum with only non-zero metric and the anti-self-dual 
five-form. The latter is given by the Freund-Rubin-type ansatz:
\bea
\label{ii}
&F_{\mu\nu\rho\kappa\lambda}=-{\sqrt{\Lambda} \over 2}
\epsilon_{\mu\nu\rho\kappa\lambda}\ , \quad 
\mu,\nu,\cdots=0,1,\cdots,4 \nn
&F_{ijkpq}=-{\sqrt{\Lambda} \over 2}
\epsilon_{ijkpq}\ , \quad 
i,j,\cdots=5,\cdots,9\ .
\eea
The vacuum has the topology of ${\rm AdS}_5\times {\rm S}^5$. 
Since ${\rm AdS}_5$ has a four dimensional Minkowski space as a 
subspace, especially on its boundary, ${\rm AdS}_5$ has the 
four dimensional Poincar\'e symmetry $ISO(1,3)$. 
${\rm S}^5$ has, of course, $SO(6)$ symmetry. 

As an extension, we can consider the solution where the 
dilaton is non-trivial but the anti-self-dual five-form is the same 
as in (\ref{ii}). Furthermore if we require the solution has 
the symmetry of $ISO(1,3)\times SO(6)$, the metric should have the 
following form:
\be
\label{iii}
ds^2=G_{\mu\nu}dx^\mu dx^\nu + g_{mn}dx^m dx^n
\ee
where $g_{mn}$ is the metric of ${\rm S}^5$ and ref.\cite{3}
\be
\label{iv}
G_{\mu\nu}dx^\mu dx^\nu
=f(y)dy^2 + y \sum_{i,j=0}^{d-1}\eta_{ij}dx^i dx^j \ .
\ee
In order to keep the symmetry of $ISO(1,3)\times SO(6)$, the 
dilaton field $\phi$ can only depend on $y$. Then by integrating 
five coordinates on ${\rm S}^5$, we obtain the effective five 
dimensional theory, which corresponds to $d=4$ and 
$\alpha={1 \over 2}$ case in (\ref{i}). We keep working with above 
dilatonic gravity as it will be easy to come to IIB supergravity 
($d=4$, $\alpha={1 \over 2}$) at any step.

 From the variation of the action (\ref{i}) with respect to 
the metric $G^{\mu\nu}$, we obtain\footnote{
The conventions of curvatures are given by
\begin{eqnarray*}
R&=&G^{\mu\nu}R_{\mu\nu} \\
R_{\mu\nu}&=& -\Gamma^\lambda_{\mu\lambda,\kappa}
+ \Gamma^\lambda_{\mu\kappa,\lambda}
- - \Gamma^\eta_{\mu\lambda}\Gamma^\lambda_{\kappa\eta}
+ \Gamma^\eta_{\mu\kappa}\Gamma^\lambda_{\lambda\eta} \\
\Gamma^\eta_{\mu\lambda}&=&{1 \over 2}G^{\eta\nu}\left(
G_{\mu\nu,\lambda} + G_{\lambda\nu,\mu} - G_{\mu\lambda,\nu} 
\right)\ .
\end{eqnarray*}
}
\be
\label{iit}
0=R_{\mu\nu}-{1 \over 2}G_{\mu\nu}R + {\Lambda \over 2}G_{\mu\nu}
- - \alpha \left(\partial_\mu\phi\partial_\nu\phi 
- -{1 \over 2}G_{\mu\nu}G^{\rho\sigma}\partial_\rho \phi
\partial_\sigma \phi \right)
\ee
and from that of dilaton $\phi$
\be
\label{iiit}
0=\partial_\mu\left(\sqrt{-G}G^{\mu\nu}\partial_\nu\phi\right)\ .
\ee
We assume that $\phi$ depends 
only on one of the coordinate, say $y\equiv x^d$ as in type IIB 
supergravity solution with the symmetry of $ISO(1,3)\times SO(6)$ 
and we also assume, as a generalization of (\ref{iv}), that 
$G_{\mu\nu}$ has the following form
\be
\label{viit}
ds_{d+1}^2=\sum_{\mu,\nu=0}^d G_{\mu\nu}dx^\mu dx^\nu
=f(y)dy^2 + y\sum_{i,j=0}^{d-1}g_{ij}dx^i dx^j
\ee
Here $g_{ij}$ is the metric in the Einstein manifold, which 
is defined by
\be
\label{vat}
r_{ij}=kg_{ij}\ .
\ee
Here $r_{ij}$ is the Ricci tensor given by $g_{ij}$ and $k$ is 
a constant, especially $k>0$ for sphere and $k=0$ for the 
flat Minkowski space and $k<0$ for hyperboloid. Such a solution
generalizes the previous solution of ref.\cite{3} (where $k=0$) 
as boundary gauge QFT lives now in four-dimensional curved spacetime.
The case of $k=1$ is especially interesting as it corresponds 
to gauge theory in de Sitter (inflationary) Universe.

The equations of motion (\ref{iit}) and (\ref{iiit}) take the 
following forms:
\bea
\label{vit}
0&=&{1 \over 2}{rf \over y}
- -{d(d-1) \over 8}{1 \over y^2} + {\lambda^2 \over 2}f 
+ {\alpha \over 2}(\phi')^2 \\
\label{viitb}
0&=&-\left(r_{ij} -{1 \over 2}rg_{ij}\right){f \over y} \nn
&& + \left\{{d-1 \over 4}{f'\over fy}
- -{(d-1)(d-4) \over 8}{1 \over y^2} 
+ {\lambda^2 \over 2}f - {\alpha \over 2}(\phi')^2\right\}g_{ij} \\
\label{viiit}
0&=&\left(\sqrt{{y^d \over f}}\phi'\right)'\ .
\eea
Here $'$ expresses the derivative with respect to $y$ and 
$r\equiv g^{ij}r_{ij}=kd$. 
Eq.(\ref{vit}) corresponds to $(\mu,\nu)=(d,d)$ in (\ref{iit}) and 
Eq.(\ref{viitb}) to $(\mu,\nu)=(i,j)$. The case of $(\mu,\nu)=(0,i)$ 
or $(i,0)$ is identically satisfied. Integrating (\ref{viiit}), 
we find 
\be
\label{ixt}
\phi'=c\sqrt{{f \over y^d}}\ .
\ee
Substituting (\ref{ixt}) into (\ref{vit}), we can solve it 
algebraically with respect to $f$:
\be
\label{xt}
f={d(d-1)  \over 4y^2
\lambda^2 \left(1 + {\alpha c^2 \over \lambda^2 y^d}
+ {kd \over \lambda^2 y}\right)}\ .
\ee
Then we find from (\ref{ixt}) and (\ref{xt}), 
\be
\label{xiiit}
\phi=c\int dy \sqrt{{d(d-1) \over 
4y^{d+2}\lambda^2 \left(1 + {\alpha c^2 \over \lambda^2 y^d}
+ {kd \over \lambda^2 y}\right)}}\ .
\ee

When $y$ is small, $f(y)$ in (\ref{xt}) behaves as
\be
\label{SG1}
f(y)\sim {d(d-1) y^{d-2} \over 4\alpha c^2}\ ,
\ee
which makes a curvature singularity at $y=0$.
The scalar curvature behaves when $y\sim 0$ as
\be
\label{Axxxviib}
R \sim \alpha c^2 y^{-d} \ .
\ee
The curvature singularity would be generated by the singular 
behavior of the dilaton $\phi$ when $y\sim 0$:
\be
\label{SG2}
\phi(y) \sim {\rm sgn}(c)\sqrt{d(d-1) \over 4\alpha} \ln y\ .
\ee
Here ${\rm sgn}(c)$ expresses the sign of $c$:
\be
\label{SG3}
{\rm sgn}(c)=\left\{\begin{array}{lll}
+1\ &\ \mbox{if}\ &\ c>0 \\
- -1\ &\ \mbox{if}\ &\ c<0 \\
\end{array}\right.\ .
\ee
The curvature singularity tells there should appear 
the $\alpha'$ correction from the string theory 
and the supergravity description would break down 
when $y\sim 0$. Conversely and hopefully, the curvature singularity 
might be apparent and vanish when we can include full string 
corrections. In any case, the solution would be valid if we 
investigate the behavior near the boundary ($y\rightarrow +\infty$).

We also note that the dilaton field behaves near the boundary 
($y\rightarrow +\infty$) as 
\be
\label{Axiiit}
\phi\sim \phi_0 - c \sqrt{{d-1 \over d\lambda^2 y^d}}
+ \cdots \ .
\ee
The term of ${\cal O}\left({1 \over y^{d \over 2}}\right)$ 
might tell that the solution given here would correspond to the 
condensation of the dimension $d$ operator, say, ${\rm tr}F^2$. 
In the usual non-(or lower-)supersymmetric QCD, it is widely believed 
that there would occur the condensation of ${\rm tr}F^2$. 
Therefore not depending on that the solution given here corresponds 
the real deformation from the ${\cal N}=4$ theory or the 
deformation of the vacuum, the solution would possibly reflect 
the structure of non-supersymmetric QCD. 

If we change the coordinate $y$ by $\rho$, which is defined by
\be
\label{xiiitb}
\rho\equiv -\int dy \sqrt{f(y) \over y}
=-\int dy \sqrt{d(d-1)  \over 4y^3
\lambda^2 \left(1 + {\alpha c^2 \over \lambda^2 y^d}
+ {kd \over \lambda^2 y}\right)}\ ,
\ee
the metric in (\ref{viit}) has the following form
\be
\label{ivtbb}
G_{\mu\nu}dx^\mu dx^\nu
=\Omega^2(\rho)\left(d\rho^2 + \sum_{i,j=0}^{d-1}g_{ij}dx^i dx^j
\right) \ .
\ee
Here $\Omega^2(\rho)$ is given by solving $y$ in (\ref{xiiitb}) 
with respect to $\rho$: $\Omega^2(\rho)=y(\rho)$.
When $\rho$ is small, $y$ is large and the structure of the 
spacetime becomes AdS asymptotically. From (\ref{xiiitb}), we find 
\be
\label{oI}
\rho={\sqrt{d(d-1)} \over \lambda y^{1 \over 2}}\left(1
+{\cal O}\left(y^{-1}\right)\right)\ .
\ee
Therefore we find
\bea
\label{oII}
&& \Omega^2(\rho)=y(\rho)={R_s^2 \over \rho^2}\left(1 
+ {\cal O}\left(\rho^2\right)\right)\nn
&& R_s\equiv {\sqrt{d(d-1)} \over \lambda }\ .
\eea
We can compare the above behavior with that of the previous 
${\rm AdS}_5\times {\rm S}^5$ solution in type IIB supergravity 
\cite{4}. The ${\rm AdS}_5$ part in the solution has the form of 
\be
\label{oIIb}
ds_{{\rm AdS}_5}^2=\left(4\pi 
g_s N\right)^{1 \over 2}\cdot{1 \over \rho^2}
\left(d\rho^2 + \sum_{i,j=0}^{d-1}\eta_{ij}dx^i dx^j
\right) \ .
\ee
Therefore we find 
\be
\label{oIIc}
R_s=\left(4\pi g_s N\right)^{1 \over 4}\ ,
\ee
where $g_s$ is the string 
coupling and $N$ is the flux of the five-form $F$ in (\ref{ii}) 
through ${\rm S}^5$, which is produced by the $N$ coincident 
D3-branes. Using the definition of $R_s$ in (\ref{oII}), the 
solution (\ref{xt}) and (\ref{xiiit}) has the following form:
\bea
\label{xtB}
f&=&{R_s^2 \over 4y^2 \left(1 + {c^2 R_s^2 \over 2d(d-1) y^d}
+ {k \over (d-1) y}\right)}\ ,\nn
\phi&=&c\int dy \sqrt{{R_s^2 \over 
4y^{d+2} \left(1 + {c^2 R_s^2 \over 2d(d-1) y^d}
+ {k \over (d-1) y}\right)}}\ .
\eea
Here we put $\alpha={1 \over 2}$ and $d=4$ in order 
to get explicitly IIB supergravity background.
On the other hand, if we change the coordinate by 
\be
\label{ivtbc}
\sigma=\int dy \sqrt{f(y)}\ ,
\ee
the metric in (\ref{viit}) has the following form
\be
\label{ivtbe}
G_{\mu\nu}dx^\mu dx^\nu
=d\sigma^2 + S(\sigma)\sum_{i,j=0}^{d-1}g_{ij}dx^i dx^j\ ,
\ee
where $S(\sigma)$ is given by solving $y$ in (\ref{ivtbc}) 
with respect to $\sigma$: $S(\sigma)=y(\sigma)$.

We now consider the case $k<0$. First let the dilaton field to be  
constant or small. Then from Eq.(\ref{xt}), when $y$ decreases 
from the positive infinity, the function $f$ increases and 
diverges at a finite value of $y$ : $y=y_0$ and after that 
the signature of the metric seems to change. This is not, however, 
real but apparent. Near $y=y_0$, the function $f(y)$ behaves as
\be
\label{ni}
f(y)\sim {f_0 \over y - y_0}\ ,
\ee
where $f_0$ is a constant. When we introduce a new coordinate $u$
by
\be
\label{nii}
y-y_0=u^2\ ,
\ee
the metric has the following form when $y\sim y_0$, 
\be
\label{niii}
ds_{d+1}^2\sim 4f_0 du^2 + y_0 
\sum_{i,j=0}^{d-1}\eta_{ij}dx^i dx^j \ .
\ee
The metric in (\ref{niii}) is regular even when $u\sim 0$ 
($y\sim y_0$) and there is no curvature singularity. 
The change of coordinates in (\ref{nii}) tells that $y$ increases 
again as $u$ increases when $u>0$. Then when we write the solution 
by the coordinate $u$, the solution connects two boundaries at 
$u=-\infty$ and $u=+\infty$. 
The structure of the spacetime, however, changes when the dilaton 
becomes large. Let us write $f(y)$ in the following form:
\be
\label{niv}
f(y)={d(d-1)  \over 4y^2\lambda^2 h(y)} \ , \quad 
h(y)\equiv 1 + {\alpha c^2 \over \lambda^2 y^d}
+ {kd \over \lambda^2 y}\ .
\ee
We now investigate the condition $h(y)$ vanishes or $f(y)$ 
diverges and changes its sign. The minimum $h_{min}$ of $h(y)$ can 
be found by the equation ${dh(y) \over dy}=0$, which can be solved 
as follows:
\be
\label{nv}
y=y_0\equiv \left(-{\alpha c^2 \over k}\right)^{1 \over d-1}
\ee
and we find
\be
\label{nvi}
h_{min}=1 + {k(d-1) \over \lambda^2}
\left(-{\alpha c^2 \over k}\right)^{-{1 \over d-1}}\ .
\ee
Therefore $h(y)$ does not vanish if $h_{min}>0$, that is
\be
\label{nvii}
c^2>c_0^2\equiv-{k \over \alpha}\left(-{\lambda^2 \over k(d-1)}
\right)^{1-d}\ .
\ee
When $c^2>c_0^2$, the solution connects the boundary at $y=\infty$ 
with the singular boundary at $y=0$ as in the $k=0$ and $k>0$ cases. 

We now consider the running of the gauge coupling. 
Usually the AdS string coupling, which is the square of the coupling 
in ${\cal N}=4$ $SU(N)$ super-Yang-Mills when $d=4$, is proportional 
to an exponential of the dilaton field $\phi$, which we assume 
in the following. From (\ref{xiiit}), when $y$ is large and $d>2$, 
we find that the dilaton field behaves as
\be
\label{ri}
\phi=\phi_0 + {c\sqrt{d(d-1)} \over 2\lambda}
\left\{-{2 y^{-{d \over 2}}\over d} 
+ {2 \over d+2}\cdot {kd \over 2\lambda^2} y^{-{d \over 2}-1}
+ \cdots \right\}\ .
\ee
Here $\cdots$ expresses the higher order terms of ${1 \over y}$. 
We now assume the gauge coupling has the following form 
\cite{5,6,7,8,9} (of course, other ways to define running gauge 
coupling might be possible)
\bea
\label{rii}
g&=&g_s\e^{2\beta\sqrt{\alpha \over d(d-1)}
\left(\phi-\phi_0\right)}\nn
&=&g_s\left\{ 1 - {2\beta c\sqrt{\alpha} \over d\lambda}
y^{-{d \over 2}} + {kd\beta c\sqrt{\alpha} \over (d+2)\lambda^3}
y^{-{d \over 2}-1} + \cdots \right\}
\eea
In case of type IIB supergravity ($\alpha={1 \over 2}$),
\be
\label{riib}
\beta=\sqrt{d(d-1) \over 2}
\ee
and  using the definition of $R_s$ in (\ref{oII}), we find 
\be
\label{riibb}
g=g_s\left\{ 1 - {cR_s \over d}
y^{-{d \over 2}} + {k c R_s^3 \over 2(d+2)(d-1)}
y^{-{d \over 2}-1} + \cdots \right\}
\ee
The next-to-leading order term is proportional to $k$ if $k\neq 0$.
This changes the renormalization group equations drastically. 
If we multiply $N^{1 \over 2}$ with $g$, we obtain the 't Hooft 
coupling $g_H=gN^{1 \over 2}$. If we define a new coordinate $U$ by 
\be
\label{riii}
y=U^2\ ,
\ee
$U$ expresses the scale on the (boundary) $d$ dimensional space
(due to holography \cite{10}).  
Following the correspondence between long-distances/high-energy 
in the AdS/CFT scheme, $U$ can be regarded as the energy scale 
of the boundary field theory. 
Then from (\ref{rii}), we obtain the following 
renormalization group equation
\be
\label{riv}
\beta(U)\equiv U{dg \over dU}= -d (g-g_s) 
- - {2kd\beta c\sqrt{\alpha} \over (d+2)\lambda^3}
\left(-{d\lambda \over 2\beta c\sqrt{\alpha} }
\right)^{{2 \over d}+1} {(g - g_s)^{{2 \over d}+1} 
\over {g_s}^{2 \over d}}
+ \cdots\ .
\ee
The leading behavior is identical with the previous 
works \cite{4,6,7,8} but the next to leading term contains 
the fractional power of $(g - g_s)$ although the square of 
$(g - g_s)$ appears for 
$k=0$ case. We should note that the qualitative behavior does not 
depend on $\beta$ which appears in the coupling (\ref{rii}).

Hence, we found that beta-function explicitly depends on the 
curvature of four-dimensional manifold. Of course, curvature 
dependence is not yet logarithmic as it happens with usual 
quantum field theories (QFTs) (perturbative consideration) 
in curved spacetime \cite{11}. The power-like running of gauge 
coupling is much stronger than in $k=0$ case. Note that previous 
discussion of power-like running includes GUTs with large internal 
dimensions \cite{12}.
In the case under investigation we get the gauge 
coupling beta-function 
as an expansion on fractional powers of gauge coupling.

We now consider the static potential between ``quark'' and 
``anti-quark''\cite{5}. We evaluate the following Nambu-Goto action 
\be
\label{rg5}
S={1 \over 2\pi}\int d\tau d\sigma \sqrt{\det\left(g^s_{\mu\nu}
\partial_\alpha x^\mu \partial_\beta x^\nu\right)}\ .
\ee
with the ``string'' metric $g^s_{\mu\nu}$, which 
could be given by multiplying a dilaton function $k(\phi)$ to 
the metric tensor in (\ref{iii}). Especially we choose $k(\phi)$ 
by
\be
\label{rg6}
k(\phi)=\e^{2\gamma
\sqrt{\alpha \over d(d-1)}\left(\phi-\phi_0\right)}
= 1 -  {2\gamma c\sqrt{\alpha} \over d\lambda y^{d \over 2}}
+ \cdots \ .
\ee
In case of type IIB supergravity, 
\be
\label{riig}
\gamma=\beta=\sqrt{d(d-1) \over 2}\ .
\ee
We consider the static configuration $x^0=\tau$, $x^1\equiv 
x=\sigma$, $x^2=x^3=\cdots=x^{d-1}=0$ and $y=y(x)$.  
We also choose the coordinates on the boundary manifold so that the 
line given by $x^0=$constant, 
$x^1\equiv x$ and $x^2=x^3=\cdots=x^{d-1}=0$ is geodesic and 
$g_{11}=1$ on the line. 
Substituting the configuration into (\ref{rg5}), we find
\be
\label{rg7}
S={T \over 2\pi}\int dx k\left(\phi(y)\right) y \sqrt{
{f(y) \over y}\left(\partial_x y\right)^2 + 1}\ .
\ee
Here $T$ is the length of the region of the definition of $\tau$.
The orbit of $y$ can be obtained by minimizing the action $S$ 
or solving the Euler-Lagrange equation 
${\delta S \over \delta y}- \partial_x\left({\delta S 
\over \delta\left(\partial_x y\right)}\right)=0$. 
The Euler-Lagrange equation tells that 
\be
\label{rg8}
E_0={k\left(\phi(y)\right) y \over \sqrt{
{f(y) \over y}\left(\partial_x y\right)^2 + 1}}
\ee
is a constant. If we assume $y$ has a finite minimum $y_0$, where 
$\partial_x y|_{y=y_0}=0$, $E_0$ is given by
\be
\label{rg9b}
E_0=k\left(\phi(y_0)\right) y_0 \ .
\ee
Introducing a parameter $t$, we parameterize $y$ by
\be
\label{rg9}
y=y_0\cosh t\ .
\ee
Then we find
\bea
\label{rg10}
{dx \over dt}&=&{y_0^{-{1 \over 2}} \over A}
\cosh^{-{3 \over 2}}t\left\{
1 + B\cosh^{-1}ty_0^{-1} + \cdots \right\}\nn
&& A\equiv{2\lambda \over \sqrt{d(d-1)}} \ ,\ \ \ 
B\equiv -{kd \over 2\lambda^2} \ .
\eea
Here we assume that $y_0$ is large enough and the orbit of the 
string does not approach to the singularity at $y=0$, where the 
supergravity description breaks down.
Taking $t\rightarrow +\infty$, we find the distance $L$ between 
"quark" and "anti-quark" is given by
\bea
\label{rg11}
L&=& {C_{3 \over 2}y_0^{-{1 \over 2}} \over A}
+ {BC_{5 \over 2} y_0^{- {3 \over 2}} \over A} + \cdots \nn
C_a&\equiv& \int_{-\infty}^\infty dt \cosh^{- a}t
= {2^{(a-1)} \Gamma\left({a \over 2}\right)^2 
\over \Gamma(a)} \ .
\eea
We should note that the large $y_0$ corresponds to small $L$. 
As one sees the next-to-leading correction to distance 
depends on the curvature of spacetime.

Eq.(\ref{rg11}) can be solved with respect to $y_0$ and we find
\be
\label{rg12}
y_0=\left({C_{3 \over 2} \over AL}\right)^2
\left\{ 1 + {2BC_{5 \over 2} \over C_{3 \over 2}}
\left({AL \over C_{3 \over 2}}\right)^2 + \cdots \right\}\ .
\ee
Using (\ref{rg8}), (\ref{rg9}) and (\ref{rg11}), we find the 
following expression for the action $S$
\bea
\label{rg13}
S&=&{T \over 2\pi}E(L) \nn
E(L)&=&\int_{-\infty}^\infty dt {dx \over dt}
{k\left(\phi(y(t))\right)^2 y(t)^2 \over 
k\left(\phi(y_0)\right) y_0} \ .
\eea
Here $E(L)$  expresses the total energy of the 
``quark''-``anti-quark'' system. 
The energy $E(L)$ in (\ref{rg13}), however, contains the divergence 
due to the self energies of the infinitely heavy ``quark'' 
and ``anti-quark''. 
The sum of their self energies can be estimated by considering the 
configuration $x^0=\tau$, $x^1=x^2=x^3=\cdots
=x^{d-1}=0$ and $y=y(\sigma)$ (note that $x_1$ vanishes here) 
and the minimum of $y$ is $y_D$ where brane would lies.
We divide the region for $y$ to two ones, 
$\infty>y>y_0$ and $y_0>y>y_D$. 
Using the parameterization of (\ref{rg9}) and identifying $t$ 
with $\sigma$ ($t=\sigma$) for the region $\infty>y>y_0$,  
we find the following expression of the sum of self energies:
\bea
\label{rg14}
E_{\rm self}&=&\int_{-\infty}^\infty dt\, 
k\left(\phi(y(t))\right)y(t)
\sqrt{ f\left(\phi(y(t))\right)
\left(\partial_t y(t)\right)^2 \over y}\nn 
&& +2\int_{y_D}^{y_0}dy k\left(\phi(y)\right)\sqrt{y f(y)}\ .
\eea
Then the finite potential between ``quark'' and ``anti-quark'' 
is given by
\bea
\label{rg15}
\lefteqn{E_{q\bar q}(L) \equiv E(L) - E_{\rm self}} \nn
&=&{1 \over A}\left({C_{3 \over 2} \over AL}\right)\left\{D_0 
+ B\left({C_{5 \over 2} D_0  \over C_{3 \over 2}}
+ D_2 \right)
\left({AL \over C_{3 \over 2}}\right)^2 + \cdots \right\} \nn
\lefteqn{D_d \equiv 2\int_0^\infty dt \cosh^{-{d+1 \over 2}}t\,
\e^{-t} + {4 \over d-1}
={2^{d-3 \over 2} \Gamma\left({d-1 \over 4}\right)^2 
\over \Gamma\left({d-1 \over 2}\right)}
 \ .}
\eea
Here we neglected the $L$ independent terms. 
Note that leading and next-to-leading term does not depend on 
the parameter $\gamma$ in (\ref{rg6}). The leading behavior is 
consistent with the previous works and attractive since 
$D_0 = -2.39628...$ but we should note that  next-to-leading 
term is linear in $L$(for $k=0$ it was cubic ), which does not depend 
on the dimension $d$. Since ${C_{5 \over 2} D_0 
\over C_{3 \over 2}} + D_2 =3.49608>0$ and $B$ is negative if 
$k$ is positive and vice versa from (\ref{rg10}). 
Therefore the linear potential term in 
(\ref{rg15}) is repulsive if $k>0$ (sphere, i.e. gauge theory 
in de Sitter Universe) and attractive if $k<0$ (hyperboloid). 

Of course, the confinement depends on the large $L$ behavior of 
the potential. When $L$ is large, however, the orbit of the string 
would approach to the curvature singularity at $y=0$, where the 
supergravity description would break down. 
Despite of this, it might be interesting to investigate the 
large $L$ behavior. Since the behavior of $f(y)$ 
and the dilaton $\phi$ when $y$ is small is given by 
(\ref{SG1}) and (\ref{SG2}), the integrand in (\ref{rg7}) 
behaves as
\bea
\label{lL1}
\lefteqn{k\left(\phi(y)\right) y \sqrt{
{f(y) \over y}\left(\partial_x y\right)^2 + 1}} \nn
&\sim& y^{{\rm sgn}(c)\sqrt{d(d-1) \over 2} +1}
\sqrt{ {d(d-1) \over 4\alpha c^2}y^{d-3}
\left(\partial_x y\right)^2 + 1} \nn
&=& \sqrt{ {d(d-1) \over 4\alpha c^2\left( 
{\rm sgn}(c)\sqrt{d(d-1) \over 2} + {d+1 \over 2} \right)^2}
\left(\partial_x \tilde U \right)^2
+ \left(\tilde U\right)^{\gamma_0} }\ .
\eea
Here
\be
\label{lL2}
\tilde U\equiv y^{{\rm sgn}(c)\sqrt{d(d-1) \over 2} 
+ {d+1 \over 2}} \ , \quad 
\gamma_0\equiv {2{\rm sgn}(c)\sqrt{d(d-1) \over 2} 
+ 2 \over {\rm sgn}(c)\sqrt{d(d-1) \over 2} 
+ {d+1 \over 2}}\ .
\ee
When $d=4$, $0<\gamma_0<2$ when $c>0$ and $\gamma<0$ when $c<0$. 
According to the analysis in \cite{GPPZ}, the orbit of the 
string goes straight to the region $y\sim 0$ when $c>0$ 
($0<\gamma_0<2$) and the 
potential becomes independent of $L$. In this case, however, 
the potential would receive the $\alpha'$ correction from 
the string theory. On the other hand, when $c<0$ ($\gamma<0$), 
there is an effective barrier which prevents the orbit of string 
from approaching into the curvature singularity and the potential 
would not receive the $\alpha'$ correction so much and the 
supergravity description would be reliable. Furthermore 
$c<0$ ($\gamma<0$) case predicts the confinement. 

We can also evaluate the potential between monopole and 
anti-monopole by using the Nambu-Goto action for $D$-string 
instead of (\ref{rg5}) (cf.ref.\cite{13}):
\be
\label{rg5m}
S={1 \over 2\pi}\int d\tau d\sigma {1 \over k(\phi)^2}
\sqrt{\det\left(g^s_{\mu\nu}
\partial_\alpha x^\mu \partial_\beta x^\nu\right)}\ .
\ee
For the static configuration $x^0=\tau$, $x^1\equiv x=\sigma$, 
$x^2=x^3=\cdots=x^{d-1}=0$ and $y=y(x)$,  we find, instead of 
(\ref{rg7})
\be
\label{rg7m}
S={T \over 2\pi}\int dx {y \over k\left(\phi(y)\right)}
\sqrt{{f(y) \over y}
\left(\partial_x y\right)^2 + 1}\ .
\ee
We should note that $k(\phi)$ is replaced by ${1 \over k(\phi)}$ 
compared with quark anti-quark case (\ref{rg7}), 
which corresponds the replacement of $\gamma\rightarrow -\gamma$. 
As the potential in (\ref{rg15}) does not depend on $\gamma$ in the 
given order, we find the monopole anti-monopole 
potential $E_{m\bar m}$ is identical with $E_{q\bar q}$ when 
$L$ is small:
\be
\label{rg15m}
E_{m\bar m}(L)=E_{q\bar q}(L)\ . 
\ee
If we consider, however, large $L$ behavior as in (\ref{lL1}), 
we find 
\bea
\label{lL1m}
\lefteqn{{y \over k\left(\phi(y)\right)} \sqrt{
{f(y) \over y}\left(\partial_x y\right)^2 + 1}} \nn
&\sim& y^{-{\rm sgn}(c)\sqrt{d(d-1) \over 2} +1}
\sqrt{ {d(d-1) \over 4\alpha c^2}y^{d-3}
\left(\partial_x y\right)^2 + 1} \nn
&=& \sqrt{ {d(d-1) \over 4\alpha c^2\left( 
- -{\rm sgn}(c)\sqrt{d(d-1) \over 2} + {d+1 \over 2} \right)^2}
\left(\partial_x \tilde U^{(m)} \right)^2
+ \left(\tilde U^{(m)}\right)^{\gamma_0^{(m)}} }\ .
\eea
Here
\be
\label{lL2m}
\tilde U^{(m)}\equiv y^{-{\rm sgn}(c)\sqrt{d(d-1) \over 2} 
+ {d+1 \over 2}} \ , \quad 
\gamma_0^{(m)}\equiv {-2{\rm sgn}(c)\sqrt{d(d-1) \over 2} 
+ 2 \over -{\rm sgn}(c)\sqrt{d(d-1) \over 2} 
+ {d+1 \over 2}}\ .
\ee
we should note that ${\rm sgn}(c)$ in (\ref{lL1}) and (\ref{lL2}) 
is replaced by $-{\rm sgn}(c)$ in (\ref{lL1m}) and (\ref{lL2m}). 
Therefore the behavior of the potential between monopole and 
anti-monopole for large $L$ is changed from that of the potential 
between quark and anti-quark, that is, monopole and anti-monopole 
would be confined for $c>0$ but would not be confined for $c<0$. 

It is not difficult to study the curvature dependence in more detail, 
for example, numerically for different choices of parameters and 
regions. Nevertheless, we do not do this as most qualitative 
features are clear.

\section{Axionic background with non-zero curvature
and non-constant dilaton}
 
Let us present now the generalization of the above IIB SG 
background with non-trivial dilaton when non-constant 
axion is included into the action. Such a study for 
the case of flat four-dimensional space has been presented earlier in 
ref.\cite{7} (for the effects of additional scalars, 
see also ref.\cite{9}).

We include the axion field $\chi$ into the action of 
type IIB supergravity ($\alpha={1 \over 2}$) in (\ref{i}),
following ref.\cite{14}
\be
\label{axi}
S=-{1 \over 16\pi G}
\int d^{d+1}x \sqrt{-G}\left(R + \lambda^2 
- - {1 \over 2} G^{\mu\nu}\partial_\mu \phi \partial_\nu \phi 
+ {1 \over 2}\e^{2\phi}
G^{\mu\nu}\partial_\mu \chi \partial_\nu \chi \right)\ .
\ee
We work in the coordinate choice (\ref{viit}) and we assume that 
the $d$-dimensional manifold is curved (\ref{vat}) and $\chi$ 
only depends on $y$. Then, instead of (\ref{vit}-\ref{viiit}), 
we obtain 
\bea
\label{axii}
0&=&{1 \over 2}{rf \over y}
- -{d(d-1) \over 8}{1 \over y^2} + {\lambda^2 \over 2}f 
+ {1 \over 4}(\phi')^2 - {1 \over 4}\e^{2\phi}(\chi')^2 \\
\label{axiii}
0&=&-\left(r_{ij} -{1 \over 2}rg_{ij}\right){f \over y} \nn
&& + \left\{{d-1 \over 4}{f'\over fy}
- -{(d-1)(d-4) \over 8}{1 \over y^2} 
+ {\lambda^2 \over 2}f \right. \nn
&& \left. - {1 \over 4}(\phi')^2 
+ {1 \over 4}\e^{2\phi}(\chi')^2 \right\}g_{ij} \\
\label{axiv}
0&=&\left(\sqrt{{y^d \over f}}\phi'\right)'
+ \sqrt{{y^d \over f}}\e^{2\phi}(\chi')^2 \\
\label{axv}
0&=&\left(\sqrt{{y^d \over f}}\e^{2\phi}\chi'\right)'\ .
\eea
Eq.(\ref{axv}) can be integrated to give
\be
\label{axvi}
\sqrt{{y^d \over f}}\e^{2\phi}\chi'=c_\chi\ .
\ee
Using (\ref{axvi}), we can delete $\chi$ in (\ref{axiv}) and obtain
\be
\label{axvii}
0=\sqrt{{y^d \over f}}\left(\sqrt{{y^d \over f}}\phi'\right)'
+ \e^{-2\phi}c_\chi^2 \ .
\ee
Eq.(\ref{axvii}) gives another integral:
\be
\label{axviii}
c_\phi={y^d \over f}(\phi')^2 - c_\chi^2 \e^{-2\phi}\ .
\ee
By using (\ref{axvi}) and (\ref{axviii}), we can delete 
$\chi'$ and $\phi'$ in (\ref{axii}) :
\be
\label{axix}
0={1 \over 2}{rf \over y}
- -{d(d-1) \over 8}{1 \over y^2} + {\lambda^2 \over 2}f 
+ {c_\phi f \over 4y^d}\ ,
\ee
which can be solved algebraically with respect to $f(y)$: 
\be
\label{axx}
f={d(d-1)  \over 4y^2
\lambda^2 \left(1 + {c_\phi \over 2\lambda^2 y^d}
+ {kd \over \lambda^2 y}\right)}\ .
\ee
The obtained metric is identical to that in (\ref{xt}), where  
the axion vanishes, if we replace $c_\phi$ in (\ref{axx}) with 
${\alpha c^2 \over 2}$. 
Therefore there appears the curvature singularity at $y=0$ again 
and the supergravity description would breaks down when $y\sim 0$.
Note that as we work with IIB SG we assume that $d=4$.

We now introduce a new coordinate $\eta$ by
\be
\label{axxi}
\eta=-\int dy \sqrt{{f \over y^d}}
=\int dy \sqrt{d(d-1)  \over 4y^{d+2}
\lambda^2 \left(1 + {c_\phi \over 2\lambda^2 y^d}
+ {kd \over \lambda^2 y}\right)}\ ,
\ee
Eqs.(\ref{axvi}) and (\ref{axviii}) can be written as follows:
\bea
\label{axxii}
c_\chi&=&\e^{2\phi}{d \chi \over d\eta} \\
\label{axxiii}
c_\phi&=&\left({d\phi \over d\eta}\right)^2 - c_\chi^2 \e^{-2\phi}\ .
\eea
Eq.(\ref{axxiii}) can be integrated to give
\be
\label{axxiv}
\e^\phi={c_\chi \over \sqrt{c_\phi}}\sinh 
\left(\sqrt{c_\phi}\left(\eta - \eta_0\right)\right)\ .
\ee
Here $\eta_0$ is a constant of the integration. Substituting (\ref{axxiv}) 
into (\ref{axxii}) and integrating it, we find
\be
\label{axxv}
\chi=\chi_0-{\sqrt{c_\phi} \over c_\chi}\coth\left(\sqrt{c_\phi} 
\left(\eta - \eta_0\right)\right)\ .
\ee
Here $\chi_0$ is a constant of the integration. Axion describes 
the running theta angle.

When $y\rightarrow +\infty$, the geometry of the spacetime 
approaches to ${\rm AdS}_5$ asymptotically. 
Then Eq.(\ref{axxi}) can be integrated perturbatively 
\be
\label{axxvi}
\eta={1 \over \lambda}\sqrt{d-1 \over d}\left({1 \over y^{d \over 2}}
- - {kd \over 2(d+2)\lambda^2 y^{{d \over 2}+1}} + \cdots \right)\ .
\ee
Here we have chosen the constant of the integration so that 
$\eta$ vanishes when $y$ goes to positive infinity. 
When $\eta$ vanishes, $\phi$ and $\chi$ behave as,
\bea
\label{axxvii}
\e^\phi&\rightarrow& -{c_\chi \over \sqrt{c_\phi}}\sinh 
\left(\eta_0\sqrt{c_\phi}\right) \nn
\chi&\rightarrow& \chi_0
+{\sqrt{c_\phi} \over c_\chi}\coth\left(\eta_0\sqrt{c_\phi}\right) \ .
\eea
We should note that $k$-dependence does not appear in $\phi$ and 
$\chi$ if we use the coordinate $\eta$ because it is hidden in
this coordinate. If we choose $\eta_0=0$, $\e^\phi\rightarrow 0$. 
Since $4\pi\e^\phi$ can be regarded as the Yang-Mills coupling 
constant and $\rho\rightarrow 0$ ($y\rightarrow +\infty$) 
corresponds to the ultraviolet fixed point from the viewpoint 
of AdS/CFT correspondence, 
the theory can be regarded as asymptotically free.

We now compare the above results with those in \cite{7} for 
$k=0$ and $d=4$. We introduce a new coordinate $r$ by
\be
\label{axxviii}
\e^{-\eta\sqrt{2c_\phi \over 3}}=\tanh \left(
{\lambda \over \sqrt{3}}(r-r_0)\right)\ .
\ee
The coordinate transformation (\ref{axxviii}) can be given in 
terms of $y$ when $k=0$ and $d=4$ 
by using (\ref{axxi}) and (\ref{axxvi}),
\be
\label{axxix}
y^2=K^4(r)\equiv \sqrt{c_\phi \over 2\lambda^2}\sinh\left(
{2\lambda \over \sqrt{3}}(r-r_0)\right)\ .
\ee
Then the metric in (\ref{viit}) for $k=0$ has the following form
\be
\label{axxx}
ds_{d+1}^2=dr^2 + K^2(r)\sum_{i,j=0}^{d-1}\eta_{ij}dx^i dx^j\ .
\ee
By using (\ref{axxviii}), the dilaton and axion fields in 
(\ref{axxiv}) and (\ref{axxv}) can be rewritten as follows
\bea
\label{axxxi}
\e^\phi&=&{c_\chi \over 2\sqrt{c_\phi}}
\left\{\left(\coth\left(
{\lambda \over \sqrt{3}}(r-r_0)\right)\right)^{\sqrt{3 \over 2}}
\e^{-\eta_0\sqrt{c_\phi}} \right. \nn
&& \left. -\left(\tanh\left({\lambda \over \sqrt{3}}(r-r_0)\right)
\right)^{\sqrt{3 \over 2}} 
\e^{-\eta_0\sqrt{c_\phi}}\right\}\nn
\chi&=&\chi_0-{\sqrt{c_\phi} \over c_\chi}\cdot
{\left(\coth\left({\lambda \over \sqrt{3}}(r-r_0)\right)
\right)^{\sqrt{3 \over 2}} + 1 \over 
\left(\coth\left({\lambda \over \sqrt{3}}(r-r_0)\right)
\right)^{\sqrt{3 \over 2}} - 1}\ .
\eea
Then the solution in \cite{7} seems to be a special case 
corresponding to $\eta_0=\chi_0=0$.

Let us consider the potential between quark and anti-quark. 
As we are interested in the case of asymptotically free theory, 
we put $\eta_0=0$ in (\ref{axxiv}) and $d=4$. Then we find 
\bea
k(\phi(y))&\equiv&\e^\phi={c_\phi R_s \over 4 c_\chi y^2}
\left(1 - {k \over 3\lambda^2y} + \cdots \right) \nn
f(y)&=&{R_s^2 \over 4y^2}\left(1- {4k \over \lambda^2 y} 
+ \cdots \right)\ .
\eea
Then in a way similar to the discussion in the second section 
where axion is not present 
instead of (\ref{rg10}) and (\ref{rg11}), we find 
\bea
\label{axxxii}
{dx \over dt}&=&{R_s \over \sqrt{2y_0}}\cosh^{-{3 \over 2}}t
\left\{1 + {2k \over \lambda^2 y_0}\left(-{1 \over \cosh t}
- - {\cosh t \over 3\left(\cosh t + 1\right)}\right) + \cdots 
\right\} \\
\label{axxxiib}
L&=&{R_s \over \sqrt{2y_0}}\left\{C_{3 \over 2} 
+ {2k \over \lambda^2 y_0}\left(-C_{5 \over 2} 
- - {E_{1 \over 2} \over 3}
\right)+ \cdots \right\} \\
E_a&\equiv & \int _{-\infty}^\infty dt {\cosh^{-a}t 
\over \cosh t + 1}\ . \nonumber
\eea
Eq.(\ref{axxxiib}) can be solved with respect to $y_0$ as follows
\be
\label{axxxiii}
y_0={1 \over 2}\left({C_{3 \over 2} \over R_s L}\right)^2
\left\{ 1 + {8k \over \lambda^2 C_{3 \over 2}}
\left(-C_{5 \over 2} - {E_{1 \over 2} \over 3}\right)
\left({C_{3 \over 2} \over R_s L}\right)^{-2}
+ \cdots \right\}\ .
\ee
Here we assume again that $y_0$ is large and $L$ is large and 
not to break the supergravity description. 
Then using (\ref{rg13}), we obtain the following expression for $E(L)$
\bea
\label{axxxiv}
E(L)&=&{R_s \over 2}\left({C_{3 \over 2} \over R_s L}\right)
\left\{C_{7 \over 2} + {k \over \lambda^2}
\left({C_{3 \over 2} \over R_s L}\right)^{-2}\left(
- -{16 \over 3}C_{9 \over 2} + {4 \over 3}E_{7 \over 2} 
\right.\right. \nn
&& \left.\left.
- -{4 \over C_{3 \over 2}}\left(C_{5 \over 2} + {E_{1 \over 2} 
\over 3}\right)\right) + \cdots \right\}\ .
\eea
Note that the integral is finite before subtraction the 
self energy of quark and anti-quark. 
We should note that the linear potential appears in the 
next-to-leading term. The coefficient $\left(
- -{16 \over 3}C_{9 \over 2} + {4 \over 3}E_{7 \over 2} 
- -{4 \over C_{3 \over 2}}\left(C_{5 \over 2} + {E_{1 \over 2} 
\over 3}\right)\right)$ of the next-to-leading 
term is negative, since $C_{3 \over 2}$, $C_{5 \over 2}$ and 
$E_{1 \over 2}$ are positive and 
$-{16 \over 3}C_{9 \over 2} + {4 \over 3}E_{7 \over 2}$ is 
negative, what can be easily found 
\bea
\label{axxxivb}
&& -{16 \over 3}C_{9 \over 2} + {4 \over 3}E_{7 \over 2} \nn
&& =-{4 \over 3}\int_{-\infty}^\infty dt \cosh^{-{9 \over 2}}t
\left(4 - {\cosh t \over \cosh t + 1}\right) \nn
&& <-{4 \over 3}\int_{-\infty}^\infty dt \cosh^{-{9 \over 2}}t
\left(4 - 1\right) \nn
&& < 0 \ .
\eea
Therefore the linear potential in the next-to-leading term 
becomes attractive if $k<0$ and repulsive if $k>0$. 
The result is consistent to the potential without axion in 
(\ref{rg15}). 
We should note that Eqs.(\ref{axxi}) and (\ref{axxiv}) tell that 
the dilaton field behaves as 
\be
\label{alL1}
\phi \sim - \sqrt{d(d-1) \over 2}\ln y\ ,
\ee
which corresponds $c<0$ case in the pure dilaton case in 
(\ref{SG2}). Since the behavior of $f(y)$ in (\ref{axx}) is 
essential identical with the pure dilaton case in (\ref{xt}), 
the supergravity description would be valid even for large $L$ 
and the confinement for quarks would be predicted (and monopoles 
would not be confined). 

We now investigate the supersymmetric background. For $k=0$ it has 
been found in ref.\cite{7}. We look for its $k$-dependent 
generalization. Since we consider the 
background where the fermion fields, that is, dilatino $\xi$ and 
gravitino $\psi_\mu$ vanish, if the variation under some of the 
supersymmetry transformations of these fermionic fields vanishes, 
the corresponding supersymmetries are preserved. The supersymmetry 
transformations of these fields are given by \cite{14}
\bea
\label{axxxv}
\delta\xi&=&-{1 \over 2}\left(\e^\phi\partial_\mu \chi
- - \partial_\mu \phi\right)\gamma^\mu \epsilon^* \ , \nn 
\delta\xi^*&=&-{1 \over 2}\left(\e^\phi\partial_\mu \chi
+ \partial_\mu \phi\right)\gamma^\mu \epsilon \ , \nn 
\delta\psi_\mu&=&\left(\nabla_\mu + {1 \over 4}\e^\phi
\partial_\mu\chi - {\lambda \over 4\sqrt{3}}\gamma_\mu\right)
\epsilon\ ,\nn
\delta\psi^*_\mu&=&\left(\nabla_\mu - {1 \over 4}\e^\phi
\partial_\mu\chi - {\lambda \over 4\sqrt{3}}\gamma_\mu\right)
\epsilon^*\ .
\eea
When substituting the solution in (\ref{axxiv}) and (\ref{axxv}) 
into $\delta\xi$ and $\delta\chi^*$, we find 
\bea
\label{axxxvi}
\delta\xi&=&-{\sqrt{c_\phi} \over 2}\left({1 - 
\cosh \left(\sqrt{c_\phi}\left(\eta - \eta_0\right)\right)
\over \sinh 
\left(\sqrt{c_\phi}\left(\eta - \eta_0\right)\right)}\right)
\gamma^\eta \epsilon^* \ , \nn 
\delta\xi^*&=&-{\sqrt{c_\phi} \over 2}\left({1 + 
\cosh \left(\sqrt{c_\phi}\left(\eta - \eta_0\right)\right)
\over \sinh 
\left(\sqrt{c_\phi}\left(\eta - \eta_0\right)\right)}\right)
\gamma^\eta \epsilon\ . 
\eea
Therefore all the supersymmetries break down in general since 
$\delta\chi$ and $\delta\chi^*$ do not vanish. 
In the limit of $c_\phi\rightarrow 0$, however, we find 
\bea
\label{axxxvib}
\delta\xi&\rightarrow& 0\ , \nn 
\delta\xi^*&=&-{1 \over \eta - \eta_0}\gamma^\eta \epsilon\ . 
\eea
Therefore there is a possibility that half of the supersymmetries 
corresponding to $\epsilon^*$ survives in this limit. It should 
be noted that, in the limit, $f(y)$ in (\ref{axx}) becomes
\be
\label{axxb}
f={d(d-1)  \over 4y^2
\lambda^2 \left(1 +  {kd \over \lambda^2 y}\right)}\ ,
\ee
which tells that the metric of the spacetime becomes nothing but 
the metric of ${\rm AdS}_5\times {\rm S}^5$ although the dilaton 
and the axion fields are non-trivial. 
Then if we choose the spinor parameter $\epsilon^*$ by using the 
Killing spinor $\zeta$ in ${\rm AdS}_5\times {\rm S}^5$
as follows\cite{14,7}
\be
\label{axxxvii}
\epsilon^*=\e^{\phi \over 4}\zeta\rightarrow c_\chi^{1 \over 4}
(\eta - \eta_0)^{1 \over 4}\zeta\ ,
\ee
$\delta\psi_\mu^*$ vanishes in the limit of $c_\phi\rightarrow 0$, 
which tells that half of the supersymmetry 
corresponding to $\epsilon^*$, in fact, survives in this limit.
This situation does not depend on $k$. Such a solution corresponds 
to some vacuum of maximally supersymmetric YM theory where 
supersymmetry is broken to ${\cal N}=2$ .
(Note that deformations of ${\cal N}=4$ super YM theory which flow 
to fixed points like in refs.\cite{15,16} may also define
running gauge coupling). In the limit of $c_\phi\rightarrow 0$, 
the solution in (\ref{axxiv}) and (\ref{axxv}) has the following form: 
\bea
\label{axxxviii}
\e^\phi&\rightarrow& c_\chi(\eta - \eta_0) \nn
\chi&\rightarrow&\chi_0 - {1 \over c_\chi(\eta - \eta_0) }\ .
\eea
Even in the limit, the theory becomes asymptotically free 
when $\eta_0=0$ since the coupling is assumed to be given by 
$\e^\phi$ vanishes in the ultraviolet limit corresponding to 
$\eta=0$. 
We should also note that the potential ($\eta_0=0$ case) between 
quark and anti-quark in (\ref{axxxiv}) is not changed in the 
leading and next-to-leading orders since $c_\phi$ is not included 
to the corresponding expression.

\section{Discussion}

In summary, we found the background of IIB supergravity with 
non-constant dilaton, non-zero curvature of four-dimensional 
space-time and with (or without) non-trivial axion. 
By assuming the coupling is given by the exponential of the 
dilaton field $\phi$, AdS/CFT 
interpretation of such a solution gives the (power-like) running 
gauge coupling and predicts its curvature dependence. In the 
presence of axion, background may have half of supersymmetries 
unbroken. In all cases, we calculated quark-antiquark 
potential and showed that the term linear on distance $L$ explicitly 
depends on the curvature. Hence, there is the possibility that 
curvature of Universe might predict the confinement.

The complete interpretation of IIB SG background via AdS/CFT 
correspondence is not yet clear. We gave the arguments that 
most probably our background corresponds to another vacuum of 
maximally supersymmetric YM theory with some non-zero VEV operator. 
However, the possibility that it may be deformation of theory to 
another less symmetric (super) YM theory is not yet completely 
ruled out. The only possibility to understand it now is to 
investigate all properties of SG background and compare it with
properties of corresponding QFT.

For example, it would be really interesting to find further 
development of such a scenario so that to present more realistic 
(logarithmic) behavior for running gauge coupling. 
Clearly, major modifications of background are necessary.
Note in that respect the recent paper \cite{17} where
it was shown that AdS orbifolds may describe the running gauge 
coupling.

\ 

\noindent
{\bf Acknowelegements.} 
We would like to thank H.B. Nielsen for discussion 
of ``two-times'' cosmology. We are also indebted to 
A. Tseytlin and A. Sugamoto for discussions and the referee o
f this work for asking the questions 
which led to significant revision of original ms.

\section*{Appendix A}

In this appendix, we point out that there are many kinds of 
Einstein manifolds which satisfy Eq.(\ref{vat}).
The Einstein equations are given by,
\be
\label{A1}
R_{\mu\nu}-{1 \over 2}g_{\mu\nu}R+{1 \over 2}\Lambda g_{\mu\nu}
= T^{\rm matter}_{\mu\nu}\ .
\ee
Here $T^{\rm matter}_{\mu\nu}$ is the energy-momentum tensor of 
the matter fields. If we consider the vacuum solution where 
$T^{\rm matter}_{\mu\nu}=0$, Eq.(\ref{A1}) can be rewritten as
\be
\label{A2}
R_{\mu\nu}={\Lambda \over 2}g_{\mu\nu}\ .
\ee
If we put $\Lambda=2k$, Eq.(\ref{A2}) is nothing but the equation 
for the Einstein manifold (\ref{vat}). The Einstein manifolds  
are not always homogeneous manifolds like flat Minkowski, 
(anti-)de Sitter space or Nariai space but they can be some 
black hole solutions like Schwarzschild black hole, 
\be
\label{schw}
ds_4^2\equiv \sum_{\mu,\nu=0}^3 g_{\mu\nu}dx^\mu dx^\nu
=-\left(1 - {r_0 \over r}\right)dt^2
+{dr^2 \over \left(1 - {r_0 \over r}\right)} + r^2 d\Omega^2\ ,
\ee
or Kerr one for $k=0$\footnote{This type of solutions for $k=0$ case 
has been considered in ref. \cite{Bur}} or Schwarzschild 
(anti-) de Sitter black hole
\be
\label{sads}
ds_4^2=-\left(1 - {\mu \over x} 
- - {2k \over 3}x^2 \right)dt^2
+ {dr^2 \over \left(1 - {\mu \over x} 
- - {2k \over 3}x^2 \right)}+ r^2 d\Omega^2\ ,
\ee
for $k\neq 0$. 
In these solutions, the curvature singularity
at $r=0$ has a form of line penetrating ${\rm AdS}_5$ and 
the horizon makes a tube surrounding the singularity. 
This configuration seems to express D-string whose boundary 
lies on the boundary of  ${\rm AdS}_5$ or possibly D3-brane. 
Especially in case of Kerr or Kerr-(anti-)de Sitter solution, 
the object corresponding to the singularity has an angular 
momentum. 

We should note that 
the dilaton depends on the geometry of the boundary manifold 
only through $k$ as in (\ref{xiiit}). Therefore the behavior 
of the running coupling or renormalization group equation is 
irrelevant with the existence of the black hole singularity.

\section*{Appendix B}
In this Appendix we present one more solution of IIB supergravity with 
two time like signatures of metric. The physical interpretation of this
solution is not quite clear as well as its dual interpretation.

It was already few times mentioned that AdS radial coordinate 
plays the role of energy coordinate via holographic correspondence. 
It is also known that in general relativity there were attempts 
to identify the energy with time flow.
Then the following interesting question appears:
Can the same sort AdS solution be re-interpreted as the one depending 
from extra time coordinates? In a sense one has then new IIB SG 
solution with a few time-like signatures. There was some 
discussion of solutions with a few times-like signatures 
in various gravitational theories.

In order to get the time dependent solution and consider a kind 
of AdS cosmology, we perform the analytic continuation 
in the solution in (\ref{xt}) and (\ref{xiiit}) with $k=0$ 
as follows:
\be
\label{iip}
c^2\rightarrow -c^2\ ,
\quad \phi_0\rightarrow 
\phi_0-{1 \over 2}\sqrt{(d-1) \over d\alpha}\ln(-1)\ .
\ee
Then we obtain the following metric and the dilaton field:
\bea
\label{iiip}
ds_{d+1}^2&=&=f(y)dy^2 + y \sum_{i,j=0}^{d-1}\eta_{ij}dx^i dx^j \\
\label{iiibp}
f&=&{d(d-1) \over 4y^2\left(
\lambda^2 - {\alpha c^2 \over y^d}\right)} \\
\label{iiicp}
\phi&=&\phi_0+{1 \over 2}\sqrt{(d-1) \over d\alpha}\ln\left\{
{-2\alpha c^2 \over \lambda^2 y^d}+1 \mp\sqrt{
\left({2\alpha c^2 \over \lambda^2 y^d}+1\right)^2 -1}\right\}\ .
\eea
We can directly check that the solution (\ref{iiip}-\ref{iiicp}) 
satisfies (\ref{iit}) and (\ref{iiit}). When 
$\lambda^2 - {\alpha c^2 \over y^d}<0$,  dilaton field $\phi$ 
is real and $f(y)$ becomes negative, which tells that $y$ can 
be regarded as another time coordinate (AdS time) besides the 
physical time coordinate in $d$-dimensional 
Minkowski space corresponding to $\eta_{ij}$ in (\ref{iiip}).
We have unusual signature of the metric with two time-like 
coordinates. Changing the coordinate $y$ by 
\be
\label{vp}
y=\left({\alpha c^2 \over \lambda^2}\right)^{1 \over d}
\sin^{2 \over d}t\ ,
\ee
we obtain the following metric and the dilaton field
\bea
\label{vip}
ds_{d+1}^2&=&-{d-1 \over d\lambda^2}dt^2 
+ \left({\alpha c^2 \over \lambda^2}\right)^{1 \over d}
\sin^{2 \over d}t\sum_{i,j=0}^{d-1}\eta_{ij}dx^i dx^j \\
\label{vibp}
\phi&=&\phi_0+{1 \over 2}\sqrt{(d-1) \over d\alpha}\ln\left(
{1\mp\cos t \over \sin t}\right)^2\ .
\eea
Note that $t=0,\pi$ corresponds to $y=0$. Therefore there is a 
curvature singularity there. This tells that $\alpha'$ expansion in 
string theories becomes unreliable and we need to exclude the region 
$t\sim 0,\pi$. Eq.(\ref{vibp}) tells that the coupling becomes 
$t$-dependent, especially in case of type IIB supergravity case 
we find 
\be
\label{tsI}
g=g_s\e^{\phi - \phi_0}=g_s
\left({1\mp\cos t \over \sin t}\right)^{\sqrt{2-{2 \over d}}}\ .
\ee
If we change the coordinate $t$ by $\tau$ as
\be
\label{tsII}
\tau=\left({d-1 \over dc\sqrt{\alpha}}\right)\int
{dt \over \sin^{1 \over d}t }\ ,
\ee
we have the metric in the following form
\bea
\label{tsIII}
ds_{d+1}^2&=&\Theta(\tau)\left(-d\tau^2 
+ \sin^{2 \over d}t\sum_{i,j=0}^{d-1}\eta_{ij}dx^i dx^j \right)\ ,
\eea
where
\be
\label{tsIV}
\Theta(\tau)\equiv 
\left({\alpha c^2 \over \lambda^2}\right)^{1 \over d}
\sin^{2 \over d}t(\tau) \ .
\ee
Note that $t$ is solved with respect to $\tau$ by using (\ref{tsII}). 
It follows from the above speculation that one can understand 
running of gauge coupling also as dependence on ``second 
time'' (AdS time). It would be interesting to understand 
if such a picture may have any physical meaning.

The conclusion drawn from such an interpretation is that AdS 
solution may contain a few times. Then the possibility of a kind of 
phase transition between these times should be considered (this is, 
of course, highly speculative). The physical time should be 
naturally defined by observer living in such a world. One possibility 
may be to introduce potential depending on angles defining the sort 
of signature of any particular dimension. Then the minimum of this 
potential may probably define the real physical time. 
In any case, the interpretation of IIB SG solution considered in 
this appendix could be understood simply as one more IIB 
SG solution.


\begin{thebibliography}{99}
\bibitem{1} J.M. Maldacena, {\sl Adv.Theor.Math.Phys.} {\bf 2} (1998) 253;
E. Witten, {\sl Adv.Theor.Math.Phys.} {\bf 2} (1998) 253;
S. Gubser, I.R. Klebanov and A.M. Polyakov, {\sl Phys.Lett.}
 {\bf B428} (1998) 105;
for an excellent review, see O. Aharony, S. Gubser, J. Maldacena, 
H. Ooguri and Y. Oz, hep-th/9905111.

\bibitem{2} I.R. Klebanov and A.A. Tseytlin, hep-th/9811035; hep-th/9812089;
 {\sl JHEP} {\bf 03} (1999) 015;
J.A. Minahan, {\sl JHEP} {\bf 01} (1999) 020; hep-th/9902125;
G. Ferretti and D. Martelli, hep-th/9811208;
A. Armoni, E. Fuchs and J. Sonnenshein, hep-th/9903090;
M. Alishahiha, A. Brandhuber and Y. Oz, hep-th/9903186;
K. Ghoroku, hep-th/9907143.

\bibitem{3} S. Nojiri and S.D. Odintsov, {\sl Phys.Lett.} {\bf B449} 
(1999) 39, hep-th/9812017.

\bibitem{4} A. Kehagias and K. Sfetsos, hep-th/9902125.

\bibitem{5} J. Maldacena, {\sl Phys.Rev.Lett.} {\bf 80} (1998) 4859.

\bibitem{6} L. Girardello, M. Petrini, M. Porrati and A. Zaffaroni,
hep-th/9903026; S. Gubser, hep-th/9902155;
S. Nojiri and S.D. Odintsov, hep-th/9904036;
R. de Mello Koch, A.Paulin-Campbell and J. Rodriques, hep-th/9903029.
K. Ghoroku, hep-th/9907143.

\bibitem{7} A. Kehagias and K. Sfetsos, hep-th/9903109.

\bibitem{8} H. Liu and A.A. Tseytlin, hep-th/9903091.

\bibitem{9} N. Constable and R.C. Myers, hep-th/9905081.

\bibitem{10} L. Susskind and E. Witten, hep-th 9805114;
A.W. Peet and J. Polchinski, hep-th/ 9809022.

\bibitem{11} I.L. Buchbinder, S.D. Odintsov and I.L: Shapiro,
 {\sl Effective Action in Quantum Gravity}, IOP Publishing,
Bristol and Philadelphia, 1992.

\bibitem{12} T.R. Taylor and G. Veneziano, {\sl Phys.Lett.} {\bf B212} 
(1988) 147; I. Antoniadis, {\sl Phys.Lett.} {\bf B246} (1990) 377;
E. Witten, {\sl Nucl.Phys.} {\bf B471} (1996) 135;
J. Lykken, {\sl Phys.Rev.} {\bf D54} (1996) 3693;
K.R. Dienes, E. Dudas and T. Gherghetta, {\sl Phys.Lett.} {\bf B436} 
 (1998) 55;
C. Bachas, {\sl JHEP} {\bf 23} (1998)9811.

\bibitem{13} D.J. Gross and H. Ooguri, {\sl Phys.Rev.} {\bf D58} 
 (1998) 106002.

\bibitem{14} G.W. Gibbons, M.B. Green and M.J. Perry, {\sl Phys.Lett.}
{\bf B370} (1996) 37; A. Tseytlin, hep-th /9612164,
{\sl Phys.Rev.Lett.} {\bf 78} (1997) 1864;
C. Chu, P.Ho and Y.Wu, hep-th/9806103.

\bibitem{15} J. Distler and F. Zamora, hep-th/9810206;
A. Karch, D. L\"ust and A. Miemic, hep-th/9901041.

\bibitem{16} D.Z. Freedman, S. Gubser, K. Pilch 
and N.P. Warner, hep-th/9904017.

\bibitem{17} K. Behrndt and D. L\"ust, hep-th/9905180.
\bibitem{Bur} A. Burinskii, hep-th 9908198.

\bibitem{GPPZ} L. Girardello, M. Petrini, M. Porrati and 
A. Zaffaroni, hep-th/9909047.

\end{thebibliography}
\end{document}